\documentclass[12pt]{article}
\usepackage{amsmath,amssymb,amsthm,amsxtra,overpic,bbm,bm,epsfig,subfigure}
\usepackage{mathrsfs}
\usepackage{graphicx}
\usepackage{color}
\usepackage{comment}
\usepackage{float}
\usepackage{cite}
\usepackage{slashed,stmaryrd}

\def\thefootnote{\fnsymbol{footnote}}

\begin{document}

\vspace{0.2cm}

\begin{center}
{\large\bf Wigner-like Parametrization of Canonical Seesaw Models}
\end{center}

\vspace{0.2cm}

\begin{center}
{\bf Shun Zhou}~\footnote{E-mail: zhoush@ihep.ac.cn}
\\
\vspace{0.2cm}
{\small Institute of High Energy Physics, Chinese Academy of Sciences, Beijing 100049, China\\
School of Physical Sciences, University of Chinese Academy of Sciences, Beijing 100049, China}
\end{center}

\vspace{1.5cm}

\begin{abstract}
In this paper, we introduce the Wigner parametrization of unitary matrices and then apply it to the full description of canonical seesaw models, which extend the Standard Model with three right-handed neutrino singlets and account simultaneously for tiny Majorana neutrino masses and the baryon number asymmetry in the Universe. In the Wigner parametrization, the strong hierarchy between the electroweak scale $\Lambda^{}_{\rm EW} \approx 10^2~{\rm GeV}$ and the seesaw scale $\Lambda^{}_{\rm SS} \approx 10^{14}~{\rm GeV}$ is generally captured by three small rotation angles $\{\vartheta^{}_1, \vartheta^{}_2, \vartheta^{}_3\} \approx {\cal O}(\Lambda^{}_{\rm EW}/\Lambda^{}_{\rm SS})$, and all the remaining parameters reside in four $3\times 3$ unitary matrices. The connection between the Wigner parametrization and those in the literature is also established.
\end{abstract}

\newpage

\def\thefootnote{\arabic{footnote}}
\setcounter{footnote}{0}

\section{Introduction} \label{sec:intro}

The extension of the Standard Model (SM) by three right-handed neutrino singlets $N^{}_{a \rm R}$ (for $a = 1, 2, 3$) with a large Majorana mass term is a simple but viable scenario to explain both tiny Majorana neutrino masses~\cite{Minkowski:1977sc} and the baryon number asymmetry in the Universe~\cite{Fukugita:1986hr}. Such a scenario, known as the canonical seesaw model, is a renormalizable field theory with the following gauge-invariant Lagrangian~\cite{Xing:2011zza}
\begin{eqnarray}
	{\cal L} = {\cal L}^{}_{\rm SM} + \overline{N^{}_{\rm R}} {\rm i}\slashed{\partial} N^{}_{\rm R} - \left[ \frac{1}{2} \overline{N^{\rm C}_{\rm R}} {\bf m}^{}_{\rm R} N^{}_{\rm R} + \overline{\ell^{}_{\rm L}} \widetilde{H} {\bf y}^{}_\nu N^{}_{\rm R} + {\rm h.c.}\right] \; ,
	\label{eq:Lag}
\end{eqnarray}
where ${\cal L}^{}_{\rm SM}$ denotes the SM Lagrangian, $\ell^{}_{\rm L}$ stands for the lepton doublet, and $\widetilde{H} \equiv {\rm i}\sigma^2 H^*$ with $H$ being the Higgs doublet, $N^{\rm C}_{\rm R} \equiv {\sf C}\overline{N^{}_{\rm R}}^{\rm T}$ with ${\sf C}$ being the charge-conjugation matrix. In addition, ${\bf m}^{}_{\rm R}$ is the Majorana mass matrix of right-handed neutrinos, while ${\bf y}^{}_\nu$ the Dirac neutrino Yukawa coupling matrix. Notice that all the $3\times 3$ matrices, such as ${\bf m}^{}_{\rm R}$ and ${\bf y}^{}_\nu$, will be denoted by letters in boldface. After the Higgs field acquires its vacuum expectation value, i.e., $\langle H \rangle = (0, v/\sqrt{2})^{\rm T}$ with $v \approx 246~{\rm GeV}$, the SM gauge symmetry is spontaneously broken and the overall neutrino mass term can be written as
\begin{eqnarray}
	{\cal L}^{}_{\rm mass} = -\frac{1}{2} \overline{\left( \begin{matrix} \nu^{}_{\rm L} & N^{\rm C}_{\rm R}\end{matrix} \right)} \left( \begin{matrix}
		{\bf 0} & {\bf m}^{}_{\rm D} \cr {\bf m}^{\rm T}_{\rm D} & {\bf m}^{}_{\rm R}
	\end{matrix} \right) \left( \begin{matrix} \nu^{\rm C}_{\rm L} \cr N^{}_{\rm R} \end{matrix} \right) + {\rm h.c.} \; ,
\label{eq:Lagmass}
\end{eqnarray}
where ${\bf m}^{}_{\rm D} = {\bf y}^{}_\nu v/\sqrt{2}$ is the Dirac neutrino mass matrix. Then, we can diagonalize the $6\times 6$ neutrino mass matrix by a $6\times 6$ unitary matrix, i.e.,
\begin{eqnarray}
	\left( \begin{matrix}
		{\bf V} & {\bf R} \cr {\bf S} & {\bf U}
	\end{matrix}\right)^\dagger \left( \begin{matrix}
	{\bf 0} & {\bf m}^{}_{\rm D} \cr {\bf m}^{\rm T}_{\rm D} & {\bf m}^{}_{\rm R} \end{matrix} \right) \left( \begin{matrix}
		{\bf V} & {\bf R} \cr {\bf S} & {\bf U}
	\end{matrix}\right)^* = \left( \begin{matrix} \widehat{\bf m} & {\bf 0} \cr {\bf 0} & \widehat{\bf M} \end{matrix} \right) \; ,
\label{eq:diag}
\end{eqnarray}
where ${\bf V}$, ${\bf R}$, ${\bf S}$ and ${\bf U}$ are all $3\times 3$ matrices, $\widehat{\bf m} \equiv {\rm diag}\{m^{}_1, m^{}_2, m^{}_3\}$ with $m^{}_i$ (for $i = 1, 2, 3$) being the masses of three light Majorana neutrinos and $\widehat{\bf M} \equiv {\rm diag}\{M^{}_1, M^{}_2, M^{}_3\}$ with $M^{}_i$ (for $i = 1, 2, 3$) being the masses of three heavy Majorana neutrinos. One can convert to the mass basis by defining the neutrino mass eigen-fields
\begin{eqnarray}
	\left( \begin{matrix}
		\widehat{\nu}^{\rm C}_{\rm L} \cr \widehat{N}^{}_{\rm R}
	\end{matrix}\right) \equiv \left( \begin{matrix}
	{\bf V} & {\bf R} \cr {\bf S} & {\bf U}
\end{matrix}\right)^{\rm T}  \left( \begin{matrix} \nu^{\rm C}_{\rm L} \cr N^{}_{\rm R} \end{matrix} \right) \; , \quad 	\left( \begin{matrix}
\widehat{\nu}^{}_{\rm L} \cr \widehat{N}^{\rm C}_{\rm R}
\end{matrix}\right) \equiv \left( \begin{matrix}
{\bf V} & {\bf R} \cr {\bf S} & {\bf U}
\end{matrix}\right)^\dagger  \left( \begin{matrix} \nu^{}_{\rm L} \cr N^{\rm C}_{\rm R} \end{matrix} \right) \; ,
\label{eq:fieldredef}
\end{eqnarray}
with $\widehat{\nu}^{}_i \equiv \widehat{\nu}^{}_{i{\rm L}} + \widehat{\nu}^{\rm C}_{i{\rm L}}$ and $\widehat{N}^{}_i \equiv \widehat{N}^{}_{i{\rm R}} + \widehat{N}^{\rm C}_{i{\rm R}}$ being the Majorana neutrino fields (i.e., $\widehat{\nu}^{}_i = \widehat{\nu}^{\rm C}_i$ and $\widehat{N}^{}_i = \widehat{N}^{\rm C}_i$ hold). Working in this basis, where the charged-lepton mass matrix is further taken to be diagonal, one can immediately write down the Lagrangian for the charged- and neutral-current interactions involving light and heavy Majorana neutrinos
\begin{eqnarray}
{\cal L}^{}_{\rm cc} &=& \frac{g}{2\sqrt{2}} \left[ \overline{l} \gamma^\mu (1 - \gamma^{}_5) {\bf V} \widehat{\nu} W^-_\mu + \overline{l} \gamma^\mu (1 - \gamma^{}_5) {\bf R} \widehat{N} W^-_\mu \right] + {\rm h.c.}\; , \label{eq:Lcc}\\
{\cal L}^{}_{\rm nc} &=& \frac{g}{4\cos \theta^{}_{\rm W}} \left[ \overline{\widehat{\nu}} \gamma^\mu (1 - \gamma^{}_5) {\bf V}^\dagger {\bf V} \widehat{\nu} Z^{}_\mu + \overline{\widehat{\nu}} \gamma^\mu (1 - \gamma^{}_5) {\bf V}^\dagger {\bf R} \widehat{N} Z^{}_\mu \right. \nonumber \\
&~& \quad \quad \quad + \left. \overline{\widehat{N}} \gamma^\mu (1 - \gamma^{}_5) {\bf R}^\dagger {\bf V} \widehat{\nu} Z^{}_\mu + \overline{\widehat{N}} \gamma^\mu (1 - \gamma^{}_5) {\bf R}^\dagger {\bf R} \widehat{N} Z^{}_\mu  \right] \; , \label{eq:Lnc}
\end{eqnarray}
where $g$ is weak gauge coupling and $\theta^{}_{\rm W}$ is the Weinberg angle~\cite{Weinberg:1967tq}. Moreover, the interaction between Majorana neutrinos and the Higgs boson $h$ is governed by
\begin{eqnarray}
{\cal L}^{}_{h} &=& -\frac{g}{4M^{}_W} h \left[\overline{\widehat{\nu}} {\bf V}^\dagger {\bf V} \widehat{\bf m} (1+\gamma^{}_5) \widehat{\nu} + \overline{\widehat{\nu}} {\bf V}^\dagger {\bf R} \widehat{\bf M} (1+\gamma^{}_5) \widehat{N} \right. \nonumber \\
&~& \quad \quad \quad \quad \left. + \overline{\widehat{N}} {\bf R}^\dagger {\bf V} \widehat{\bf m} (1+\gamma^{}_5) \widehat{\nu} + \overline{\widehat{N}} {\bf R}^\dagger {\bf R} \widehat{\bf M} (1+\gamma^{}_5) \widehat{N} \right] + {\rm h.c.} \; , \label{eq:Lh}
\end{eqnarray}
where $M^{}_W = gv/2$ is the mass of weak gauge bosons $W^\pm$. In deriving the above interaction terms, we have made use of the unitarity conditions ${\bf V}^\dagger {\bf V} + {\bf S}^\dagger {\bf S} = {\bf R}^\dagger {\bf R} + {\bf U}^\dagger {\bf U} = {\bf 1}$ and ${\bf V}^\dagger {\bf R} + {\bf S}^\dagger {\bf U} = {\bf 0}$. It should be emphasized that only the $3\times 3$ matrices ${\bf V}$ and ${\bf R}$ are relevant for the interactions of light and heavy Majorana neutrinos in Eqs.~(\ref{eq:Lcc})-(\ref{eq:Lh}), although the $6\times 6$ unitary matrix has been introduced to diagonalize the overall neutrino mass matrix as shown in Eq.~(\ref{eq:diag}). Therefore, some parameters in the $6\times 6$ unitary matrix are unphysical.

In this work, we advocate a novel parametrization of the $6\times 6$ unitary matrix and apply it to the canonical seesaw model. Although it was proposed in 1968 by Eugene Wigner in the paper entitled ``On a Generalization of Euler's Angles" in a very different context~\cite{Wigner:1968}, it is reasonable to call it Wigner parametrization. The basic idea is to parametrize the $6\times 6$ unitary matrix as
\begin{eqnarray}
{\cal U} \equiv \left( \begin{matrix} {\bf V} & {\bf R} \cr {\bf S} & {\bf U}\end{matrix} \right) = \left( \begin{matrix} {\bf u}^{}_1 & {\bf 0} \cr {\bf 0} & {\bf u}^{}_2 \end{matrix} \right) \cdot \left( \begin{matrix} \widehat{\bf c} & \widehat{\bf s} \cr -\widehat{\bf s} & \widehat{\bf c} \end{matrix} \right) \cdot \left( \begin{matrix} {\bf v}^{}_1 & {\bf 0} \cr {\bf 0} & {\bf v}^{}_2 \end{matrix} \right) \; ,
\label{eq:wigner}
\end{eqnarray}
where ${\bf u}^{}_1$, ${\bf u}^{}_2$, ${\bf v}^{}_1$, and ${\bf v}^{}_2$ are all three-dimensional unitary matrices, $\widehat{\bf c} = {\rm diag}\{c^{}_1, c^{}_2, c^{}_3\}$ and $\widehat{\bf s} = {\rm diag}\{s^{}_1, s^{}_2, s^{}_3\}$ are diagonal matrices with $c^{}_i \equiv \cos \vartheta^{}_i$ and $s^{}_i \equiv \sin \vartheta^{}_i$ (for $i = 1, 2, 3$) for three rotation angles $\vartheta^{}_i \in (0, \pi/2)$. When applied to the canonical seesaw model, the Wigner parametrization becomes physically attractive in the sense that the strong hierarchy between the electroweak scale $\Lambda^{}_{\rm EW} \approx 10^2~{\rm GeV}$ and the seesaw scale $\Lambda^{}_{\rm SS} \approx 10^{14}~{\rm GeV}$ manifests itself in three small rotation angles $\{\vartheta^{}_1, \vartheta^{}_2, \vartheta^{}_3\} \approx {\cal O}(\Lambda^{}_{\rm EW}/\Lambda^{}_{\rm SS})$. Meanwhile all the other rotation angles and CP-violating phases are contained in four $3\times 3$ unitary matrices.

The remaining part of this paper is structured as follows. In Sec.~\ref{sec:wigner}, we explain the basic idea of Wigner parametrization and demonstrate its validity in more general cases. Then, we shall apply in Sec.~\ref{sec:applic} this parametrization to the canonical seesaw model and establish its connection to other parametrizations in the literature. Our main results and conclusions are finally summarized in Sec.~\ref{sec:summ}.

\section{Wigner Parametrization}\label{sec:wigner}

In Ref.~\cite{Wigner:1968}, Wigner first started with a simple parametrization of two-dimensional unitary matrices, which will also be considered here as an illustrative example. As is well known, a $2\times 2$ unitary matrix can be cast into
\begin{eqnarray}
{\cal U}^{}_{2} = \left( \begin{matrix} u^{}_1 & 0 \cr 0 & u^{}_2 \end{matrix} \right) \cdot \left( \begin{matrix} c & s \cr -s & c \end{matrix} \right) \cdot \left( \begin{matrix} v^{}_1 & 0 \cr 0 & v^{}_2 \end{matrix} \right) \; ,
\label{eq:2dU}
\end{eqnarray}
where $c \equiv \cos\theta$ and $s \equiv \sin\theta$ have been defined as usual, $\{u^{}_1, u^{}_2, v^{}_1, v^{}_2\}$ are one-dimensional unitary matrices, namely, complex numbers of modulus one. Since an arbitrary $2\times 2$ unitary matrix should possess four real parameters,
there is one redundant phase on the right-hand side of Eq.~(\ref{eq:2dU}). Without loss of generality, we can choose $v^{}_2 = 1$. Then, Wigner generalized this parametrization to the case of $(n+m)\times(n+m)$ unitary matrices, i.e.,\footnote{This is known as the CS decomposition of unitary matrices in mathematical literature~\cite{Horn}, where ``C" and ``S" refer to ``cosine" and ``sine", respectively. The CS decomposition is also called the generalized singular value decomposition. Wigner didn't cite any references regarding this decomposition in his paper~\cite{Wigner:1968}, so it is likely that Wigner invented it and then applied it to nuclear physics for the first time. See, e.g., Ref.~\cite{history}, for a historical survey on the contributions from mathematicians to the CS decomposition.}
\begin{eqnarray}
{\cal U}^{}_{n+m} = \left( \begin{matrix} {\bf u}^{}_1 & {\bf 0} \cr {\bf 0} & {\bf u}^{}_2 \end{matrix} \right) \cdot \left( \begin{matrix} \widehat{\bf c} & \widehat{\bf s} \cr -\widehat{\bf s}^{\rm T} & \widehat{\bf c}^\prime \end{matrix} \right) \cdot \left( \begin{matrix} {\bf v}^{}_1 & {\bf 0} \cr {\bf 0} & {\bf v}^{}_2 \end{matrix} \right) \; ,
\label{eq:mnU}
\end{eqnarray}
where ${\bf u}^{}_1$ and ${\bf v}^{}_1$ are $n\times n$ unitary matrices, ${\bf u}^{}_2$ and ${\bf v}^{}_2$ are $m\times m$ unitary matrices. In addition, $\widehat{\bf c}$ and $\widehat{\bf c}^\prime$ are real diagonal matrices; the last $n-m$ diagonal elements of $\widehat{\bf c}$ are one (for $n>m$) while the first $m$ diagonal elements are identical with those of $\widehat{\bf c}^\prime$. The $n\times m$ matrix $\widehat{\bf s}$ is called diagonal if its sub-matrix formed by the first $m$ rows (for $n > m$) is diagonal and the last $n-m$ rows are zero, so its transpose $\widehat{\bf s}^{\rm T}$ is also diagonal. The following conditions should be satisfied
\begin{eqnarray}
\widehat{\bf c}^2 + \widehat{\bf s}\widehat{\bf s}^{\rm T} = {\bf 1}^{}_n \; , \qquad \widehat{\bf c}^{\prime 2} + \widehat{\bf s}^{\rm T}\widehat{\bf s} = {\bf 1}^{}_m \; ,
\label{eq:mnid}
\end{eqnarray}
where ${\bf 1}^{}_n$ and ${\bf 1}^{}_m$ refer to the $n$- and $m$-dimensional identity matrix, respectively. In the case of $n = m$, the diagonal matrix takes its ordinary definition.

The original Wigner parametrization of $(n+m)\times (n+m)$ unitary matrices, as in Eq.~(\ref{eq:mnU}), contains more than necessary parameters. The $(n+m)$-dimensional unitary matrix should involve $(n+m)^2$ real parameters, but the number of real parameters on the right-hand side of Eq.~(\ref{eq:mnU}) is actually $2 n^2 + 2 m^2 + m$, implying that $(n-m)^2 + m$ parameters are redundant. For this reason, we reexamine Wigner's derivation in this section in order to sort out how to get rid of those redundant parameters.

First, we partition the $(n+m)$-dimensional unitary matrix ${\cal U}^{}_{n+m}$ into four blocks, whose dimensions are $n\times n$, $n\times m$, $m\times n$ and $m\times m$, respectively. For clarity, we shall take $n> m$. The four blocks are as follows
\begin{eqnarray}
{\cal U}^{}_{n+m} = \left( \begin{matrix} {\bf U}^{}_{nn} & {\bf U}^{}_{nm} \cr {\bf U}^{}_{mn} & {\bf U}^{}_{mm}\end{matrix} \right) \; ,
\end{eqnarray}
where the unitarity condition ${\cal U}^\dagger_{n+m} {\cal U}^{}_{n+m} = {\bf 1}^{}_{n+m}$ leads to
\begin{eqnarray}
{\bf U}^\dagger_{nn} {\bf U}^{}_{nn} + {\bf U}^\dagger_{mn} {\bf U}^{}_{mn} &=& {\bf 1}^{}_{n} \; , \label{eq:1nn}\\
{\bf U}^\dagger_{nn} {\bf U}^{}_{nm} + {\bf U}^\dagger_{mn} {\bf U}^{}_{mm} &=& {\bf 0}^{}_{nm} \; , \label{eq:0nm}\\
{\bf U}^\dagger_{nm} {\bf U}^{}_{nn} + {\bf U}^\dagger_{mm} {\bf U}^{}_{mn} &=& {\bf 0}^{}_{mn} \; . \label{eq:0mn}\\
{\bf U}^\dagger_{nm} {\bf U}^{}_{nm} + {\bf U}^\dagger_{mm} {\bf U}^{}_{mm}  &=& {\bf 1}^{}_{m} \; , \label{eq:1mm}
\end{eqnarray}
where the identity in Eq.~(\ref{eq:0mn}) is just the Hermitian conjugate of that in Eq.~(\ref{eq:0nm}) and thus brings in no more information. As ${\bf U}^\dagger_{mn}{\bf U}^{}_{mn}$ is an $n\times n$ Hermitian matrix and of rank $m$ at most, it can be diagonalized by a unitary matrix ${\bf v}^{}_1$ of dimension $n$ via
\begin{eqnarray}
{\bf U}^\dagger_{mn}{\bf U}^{}_{mn} = {\bf v}^\dagger_1 {\bf s}^2 {\bf v}^{}_1 \; ,
\label{eq:v1}
\end{eqnarray}
where ${\bf s}^2 \equiv {\rm diag}\{s^2_1, s^2_2, \cdots, s^2_m\} \oplus {\bf 0}^{}_{n-m}$ is an $n$-dimensional diagonal matrix with at most $m$ positive eigenvalues $s^2_i \equiv \sin^2\vartheta^{}_i$ and $\vartheta^{}_i \in (0,\pi/2)$. Since ${\bf U}^\dagger_{mn} {\bf U}^{}_{mn}$ is at most of rank $m$ and we do assume its rank to be $m$ in subsequent discussions, ${\bf v}^{}_1$ cannot be a general $n\times n$ unitary matrix. The $n-m$ zero eigenvalues of ${\bf s}^2$ indicate that the unitary transformation with an arbitrary $(n-m)\times (n-m)$ unitary matrix with $(n-m)^2$ real parameters in the null space corresponding to zero eigenvalues doesn't change Eq.~(\ref{eq:v1}). Moreover, the common phases of the first $m$ rows of ${\bf v}^{}_1$ cancel out on the right-hand side of Eq.~(\ref{eq:v1}). As a consequence, the number of relevant real parameters in ${\bf v}^{}_1$ is actually $n^2 - (n-m)^2 - m$. This parameter counting explains the existence of redundant parameters in the Wigner parametrization in Eq.~(\ref{eq:mnU}) and indicates how to eliminate them.

From Eq.~(\ref{eq:1nn}) and Eq.~(\ref{eq:v1}), we can observe
\begin{eqnarray}
{\bf U}^\dagger_{nn} {\bf U}^{}_{nn} = {\bf 1}^{}_n - {\bf v}^\dagger_1 {\bf s}^2 {\bf v}^{}_1 = {\bf v}^\dagger_1 {\bf c}^2 {\bf v}^{}_1 \; ,
\label{eq:u1}
\end{eqnarray}
where ${\bf c}^2 \equiv {\rm diag}\{c^2_1, c^2_2, \cdots, c^2_m\} \oplus {\bf 1}^{}_{n-m}$ with $c^2_i \equiv \cos^2\vartheta^{}_i$ is an $n\times n$ diagonal matrix and $c^2_i + s^2_i = 1$ is automatically satisfied for $i = 1, 2, \cdots, m$ by definition. It is obvious that one can introduce $\widehat{\bf c} \equiv {\rm diag}\{c^{}_1, c^{}_2, \cdots, c^{}_m\} \oplus {\bf 1}^{}_{n-m}$ such that ${\bf c}^2 = \widehat{\bf c}^2$ and $\widehat{\bf c} {\bf v}^{}_1$ is nonsingular. Therefore, we have
\begin{eqnarray}
{\bf U}^{}_{nn} = {\bf u}^{}_1 \widehat{\bf c} {\bf v}^{}_1 \; ,
\label{eq:Unn}
\end{eqnarray}
with ${\bf u}^{}_1$ being an $n\times n$ unitary matrix.

Then, we shall reconstruct the block matrix ${\bf U}^{}_{mn}$ from Eq.~(\ref{eq:v1}). To this end, we introduce an $m\times n$ diagonal matrix $\widehat{\bf s}^{\rm T} = ({\rm diag}\{s^{}_1, s^{}_2, \cdots, s^{}_m\}, {\bf 0}^{}_{m,n-m})$, where the square matrix formed by first $m$ columns is diagonal and the rest $n-m$ columns are all zero. This matrix is in fact the transpose of the $n\times m$ diagonal matrix $\widehat{\bf s} \equiv ({\rm diag}\{s^{}_1, s^{}_2, \cdots, s^{}_m\}, {\bf 0}^{}_{m,n-m})^{\rm T}$, and we have $\widehat{\bf s}^{\rm T} = \widehat{\bf s}$ for $n = m$. Notice that $\widehat{\bf s} \widehat{\bf s}^{\rm T} = {\bf s}^2$, so one can replace ${\bf s}^2$ on the right-hand side of Eq.~(\ref{eq:v1}) by $\widehat{\bf s} \widehat{\bf s}^{\rm T}$, namely,
\begin{eqnarray}
{\bf U}^\dagger_{mn}{\bf U}^{}_{mn} = {\bf v}^\dagger_1 \widehat{\bf s} \widehat{\bf s}^{\rm T} {\bf v}^{}_1 = \left(\widehat{\bf s}^{\rm T} {\bf v}^{}_1\right)^\dagger \cdot (\widehat{\bf s}^{\rm T} {\bf v}^{}_1) \; .
\end{eqnarray}
It is straightforward to prove that both ${\bf U}^{}_{mn}$ and $\widehat{\bf s}^{\rm T} {\bf v}^{}_1$ are of rank $m$, and thus there exists a unitary matrix ${\bf u}^{}_2$ of dimension $m$ to relate one to another, i.e.,
\begin{eqnarray}
{\bf U}^{}_{mn} = - {\bf u}^{}_2 \widehat{\bf s}^{\rm T} {\bf v}^{}_1 \; ,
\label{eq:Umn}
\end{eqnarray}
where the minus sign on the right-hand side has been chosen simply as our convention. This is analogous to the two-dimensional example in Eq.~(\ref{eq:2dU}).

Finally, we study the other two blocks ${\bf U}^{}_{nm}$ and ${\bf U}^{}_{mm}$ in a similar way. Starting with Eq.~(\ref{eq:1mm}), one can diagonalize the $m$-dimensional Hermitian matrix ${\bf U}^\dagger_{nm} {\bf U}^{}_{nm}$ via
\begin{eqnarray}
{\bf U}^\dagger_{nm} {\bf U}^{}_{nm} = {\bf v}^\dagger_2 {\bf s}^{\prime 2} {\bf v}^{}_2 \; ,
\label{eq:v2}
\end{eqnarray}
where ${\bf v}^{}_2$ stands for an $m$-dimensional unitary matrix and ${\bf s}^{\prime 2} \equiv {\rm diag}\{s^{\prime 2}_1, s^{\prime 2}_2, \cdots, {s}^{\prime 2}_m\}$ is an $m$-dimensional diagonal matrix with positive eigenvalues ${s}^{\prime 2}_i \equiv \sin^2 \vartheta^{\prime}_i$ for $i = 1, 2, \cdots, m$. It is worth mentioning that the common phases of $m$ rows in ${\bf v}^{}_2$ cancel out in Eq.~(\ref{eq:v2}), as for ${\bf v}^{}_1$ in Eq.~(\ref{eq:v1}), so one can remove those $m$ phases either in ${\bf v}^{}_1$ (as we have done) or in ${\bf v}^{}_2$, but not both. Otherwise, it will be problematic to fulfill the orthogonality condition in Eq.~(\ref{eq:0nm}) or equivalently in Eq.~(\ref{eq:0mn}).

To further pin down ${\bf U}^{}_{nm}$, we define $\widehat{\bf s}^\prime \equiv ({\rm diag}\{s^\prime_1, s^\prime_2, \cdots, s^\prime_m\}, {\bf 0}^{}_{m,n-m})^{\rm T}$, which is an $n\times m$ diagonal matrix. In analogy with the determination of ${\bf U}^{}_{mn}$, it is easy to get
\begin{eqnarray}
{\bf U}^{}_{nm} = {\bf u}^\prime_1 \widehat{\bf s}^\prime {\bf v}^{}_2 \; ,
\label{eq:Unm}
\end{eqnarray}
where ${\bf u}^\prime_1$ is an $n\times n$ unitary matrix. On the other hand, as in the case of ${\bf U}^{}_{nn}$, we have
\begin{eqnarray}
{\bf U}^\dagger_{mm} {\bf U}^{}_{mm} = {\bf v}^\dagger_2 (1 - {\bf s}^{\prime 2}) {\bf v}^{}_2 = {\bf v}^\dagger_2 {\bf c}^{\prime 2} {\bf v}^{}_2 \; ,
\label{eq:Umm}
\end{eqnarray}
with ${\bf c}^{\prime 2} = {\rm diag}\{{c}^{\prime 2}_1, {c}^{\prime 2}_2, \cdots, {c}^{\prime 2}_m\}$ and ${c}^{\prime 2}_i = \cos^2{\vartheta}^{\prime}_i$ for $i = 1, 2, \cdots, m$. Similarly, after defining $\widehat{\bf c}^\prime \equiv {\rm diag}\{c^\prime_1, c^\prime_2, \cdots, c^\prime_m\}$ and noticing ${\bf c}^{\prime 2} = \widehat{\bf c}^{\prime 2}$, one arrives at
\begin{eqnarray}
{\bf U}^{}_{mm} = {\bf u}^\prime_2 \widehat{\bf c}^\prime {\bf v}^{}_2 \; ,
\label{eq:Umm2}
\end{eqnarray}
with an $m$-dimensional unitary matrix ${\bf u}^\prime_2$. With the help of Eq.~(\ref{eq:0nm}), it is now ready to relate $\{{\bf u}^\prime_1, {\bf u}^\prime_2\}$ and $\{\widehat{\bf c}^\prime, \widehat{\bf s}^\prime\}$ to the previously known matrices $\{{\bf u}^{}_1, {\bf u}^{}_2\}$ and $\{\widehat{\bf c}, \widehat{\bf s}\}$. More explicitly, inserting Eqs.~(\ref{eq:Unn}), (\ref{eq:Umn}), (\ref{eq:Unm}) and (\ref{eq:Umm2}) into Eq.~(\ref{eq:0nm}) and following the detailed arguments given in Ref.~\cite{Wigner:1968}, we can obtain ${\bf u}^\prime_1 = {\bf u}^{}_1$, ${\bf u}^\prime_2 = {\bf u}^{}_2$ and $\widehat{\bf s}^\prime = \widehat{\bf s}$ and that the diagonal matrix formed by first $m$ rows of $\widehat{\bf c}$ is identical to $\widehat{\bf c}^\prime$. Putting all together, one has proved the validity of Wigner parametrization in Eq.~(\ref{eq:mnU}). Only the case of $n = m = 3$ will be of our interest in this work, but the results for $n > m$ are useful for the minimal seesaw model with two right-handed neutrinos (i.e., $n=3$ and $m=2$)\cite{Kleppe:1995zz, Borstnik:1999wu}.

\section{Application to Seesaw Models}\label{sec:applic}

In this section, we proceed to apply the Wigner parametrization to the canonical seesaw model. Adopting the Wigner parametrization in Eq.~(\ref{eq:wigner}) for the $6\times 6$ unitary matrix ${\cal U}$, we diagonalize the overall neutrino mass matrix via ${\cal U}^\dagger {\cal M} {\cal U}^* = \widehat{\cal M}$, but now rewrite the results in a more suggestive way
\begin{eqnarray}
\left( \begin{matrix} \widehat{\bf c} & -\widehat{\bf s} \cr \widehat{\bf s} & \widehat{\bf c}\end{matrix} \right) \cdot \left( \begin{matrix} {\bf 0} & {\bf u}^\dagger_1 {\bf m}^{}_{\rm D} {\bf u}^*_2 \cr {\bf u}^\dagger_2 {\bf m}^{\rm T}_{\rm D} {\bf u}^*_1 & {\bf u}^\dagger_2 {\bf m}^{}_{\rm R} {\bf u}^*_2 \end{matrix}\right) \cdot \left( \begin{matrix} \widehat{\bf c} & \widehat{\bf s} \cr -\widehat{\bf s} & \widehat{\bf c}\end{matrix} \right) = \left( \begin{matrix} {\bf v}^{}_1 \widehat{\bf m} {\bf v}^{\rm T}_1 & {\bf 0} \cr {\bf 0} & {\bf v}^{}_2 \widehat{\bf M} {\bf v}^{\rm T}_2\end{matrix} \right) \; .
\label{eq:key}
\end{eqnarray}
Some helpful comments on the above equation are in order. First, we have separated two sets of unitary matrices $\{{\bf u}^{}_1, {\bf u}^{}_2\}$ and $\{{\bf v}^{}_1, {\bf v}^{}_2\}$ into the left- and right-hand side, respectively. In this way, it becomes evident that the role played by the block-diagonal orthogonal matrix (i.e., the CS matrix) is to block-diagonalize the modified version of the overall neutrino mass matrix. Second, since only a very special form of the $6\times 6$ symmetric matrix can be block-diagonalized by the CS matrix, ${\bf u}^{}_1$ and ${\bf u}^{}_2$ are supposed to transform the sub-matrices ${\bf m}^{}_{\rm D}$ and ${\bf m}^{}_{\rm R}$ into those desired forms. Third, comparing between the parametrizations of ${\cal U}$ in Eq.~(\ref{eq:wigner}), we can find
\begin{eqnarray}
{\bf V} = {\bf u}^{}_1 \cdot \widehat{\bf c} \cdot {\bf v}^{}_1 \; , \qquad {\bf R} = {\bf u}^{}_1 \cdot \widehat{\bf s} \cdot {\bf v}^{}_2 \; ,
\label{eq:VR}
\end{eqnarray}
implying that the Wigner parametrization is essentially the singular value decomposition of ${\bf V}$ and ${\bf R}$ with their singular values in $\widehat{\bf c}$ and $\widehat{\bf s}$, respectively.

Since it is quite straightforward to further parametrize the $3\times 3$ unitary matrices ${\bf u}^{}_i$ and ${\bf v}^{}_i$ (for $i = 1, 2$) by rotation angles and CP-violating phases, we shall demonstrate how to find out them together with three rotation angles $\{\vartheta^{}_1, \vartheta^{}_2, \vartheta^{}_3\}$ in the CS matrix for a given set of ${\bf m}^{}_{\rm D}$ and ${\bf m}^{}_{\rm R}$. From Eq.~(\ref{eq:key}), we can obtain
\begin{eqnarray}
{\bf v}^{}_1 \widehat{\bf m} {\bf v}^{\rm T}_1 &=& -\widehat{\bf s} \widetilde{\bf m}^{\rm T}_{\rm D} \widehat{\bf c} -\widehat{\bf c} \widetilde{\bf m}^{}_{\rm D} \widehat{\bf s} + \widehat{\bf s} \widetilde{\bf m}^{}_{\rm R} \widehat{\bf s} \; , \label{eq:mLeff}\\
{\bf 0} &=& -\widehat{\bf s} \widetilde{\bf m}^{\rm T}_{\rm D} \widehat{\bf s} + \widehat{\bf c} \widetilde{\bf m}^{}_{\rm D} \widehat{\bf c} - \widehat{\bf s} \widetilde{\bf m}^{}_{\rm R} \widehat{\bf c} \; , \label{eq:offdiag12}\\
{\bf 0} &=& +\widehat{\bf c} \widetilde{\bf m}^{\rm T}_{\rm D} \widehat{\bf c} - \widehat{\bf s} \widetilde{\bf m}^{}_{\rm D} \widehat{\bf s} - \widehat{\bf c} \widetilde{\bf m}^{}_{\rm R} \widehat{\bf s} \; , \label{eq:offdiag21}\\
{\bf v}^{}_2 \widehat{\bf M} {\bf v}^{\rm T}_2 &=& +\widehat{\bf c} \widetilde{\bf m}^{\rm T}_{\rm D} \widehat{\bf s} + \widehat{\bf s} \widetilde{\bf m}^{}_{\rm D} \widehat{\bf c} + \widehat{\bf c} \widetilde{\bf m}^{}_{\rm R} \widehat{\bf c} \; , \label{eq:mReff}
\end{eqnarray}
where $\widetilde{\bf m}^{}_{\rm D} \equiv {\bf u}^\dagger_1 {\bf m}^{}_{\rm D} {\bf u}^*_2$ and $\widetilde{\bf m}^{}_{\rm R} \equiv {\bf u}^\dagger_2 {\bf m}^{}_{\rm R} {\bf u}^*_2$ have been defined.\footnote{We notice that a special case of $\widetilde{\bf m}^{}_{\rm D} = {\rm diag}\{d^{}_1, d^{}_2, d^{}_3\}$ and $\widetilde{\bf m}^{}_{\rm R} = {\rm diag}\{r, r, r\}$ with all positive eigenvalues has previously been discussed in Ref.~\cite{Cheng:1980tp}, where it is easy to derive $\tan2\vartheta^{}_i = 2d^{}_i/r$ and thus ${\bf v}^{}_1 = {\rm i}{\bf 1}$, ${\bf v}^{}_2 = {\bf 1}$, $m^{}_i =(\sqrt{r^2 + 4d^2_i}-r)/2$ and $M^{}_i = (\sqrt{r^2 + 4d^2_i}+r)/2$. Therefore, the flavor mixing matrices appearing in the neutrino interactions are ${\bf V} = {\rm i}{\bf u}^{}_1 \cdot \widehat{\bf c}$ and ${\bf R} = {\bf u}^{}_1 \cdot \widehat{\bf s}$. In the most general case, however, it is not expected that both $\widetilde{\bf m}^{}_{\rm D}$ and $\widetilde{\bf m}^{}_{\rm R}$ can be diagonal simultaneously.} A number of interesting observations can be made from these relations:
\begin{itemize}
\item If we multiply Eq.~(\ref{eq:offdiag12}) by $\widehat{\bf c}^{-1} \widehat{\bf s}$ from the right and add it to Eq.~(\ref{eq:mLeff}) on both sides, then one can get a simple formula
    \begin{eqnarray}
    {\bf v}^{}_1 \widehat{\bf m} {\bf v}^{\rm T}_1 = -\widehat{\bf s} \widetilde{\bf m}^{\rm T}_{\rm D} \widehat{\bf c} \left( {\bf 1} + \widehat{\bf s}^2/\widehat{\bf c}^2  \right) = -\widehat{\bf s} \widetilde{\bf m}^{\rm T}_{\rm D} \widehat{\bf c}^{-1} \; .
    \label{eq:mLeffsim}
    \end{eqnarray}
    In a similar way, multiplying Eq.~(\ref{eq:offdiag21}) by $\widehat{\bf s}^{-1} \widehat{\bf c}$ from the right and add it to Eq.~(\ref{eq:mReff}) on both sides, we have
    \begin{eqnarray}
    {\bf v}^{}_2 \widehat{\bf M} {\bf v}^{\rm T}_2 = +\widehat{\bf c} \widetilde{\bf m}^{\rm T}_{\rm D} \widehat{\bf s} \left( {\bf 1} + \widehat{\bf c}^2/\widehat{\bf s}^2  \right) = +\widehat{\bf c} \widetilde{\bf m}^{\rm T}_{\rm D} \widehat{\bf s}^{-1} \; .
    \label{eq:mReffsim}
    \end{eqnarray}
    It is worthwhile to stress that the effective mass matrices for light and heavy Majorana neutrinos in Eqs.~(\ref{eq:mLeffsim}) and (\ref{eq:mReffsim}) after the block diagonalization have been simplified in a remarkable way. Moreover, both effective neutrino mass matrices are in fact symmetric, so are $\widehat{\bf s} \widetilde{\bf m}^{\rm T}_{\rm D} \widehat{\bf c}^{-1}$ and $\widehat{\bf c} \widetilde{\bf m}^{\rm T}_{\rm D} \widehat{\bf s}^{-1}$. This observation leads to $\widehat{\bf c}\widehat{\bf s} \widetilde{\bf m}^{\rm T}_{\rm D} = \widetilde{\bf m}^{}_{\rm D} \widehat{\bf c} \widehat{\bf s}$, or equivalently,
    \begin{eqnarray}
    s^{}_i c^{}_i \left(\widetilde{\bf m}^{}_{\rm D}\right)^{}_{ji} = s^{}_j c^{}_j \left(\widetilde{\bf m}^{}_{\rm D}\right)^{}_{ij} \;.
    \label{eq:DijDji}
    \end{eqnarray}
    This identity imposes very restrictive constraints on $\widetilde{\bf m}^{}_{\rm D}$, and thus on ${\bf u}^{}_1$ and ${\bf u}^{}_2$, which we aim to determine.

\item As the identity in Eq.~(\ref{eq:offdiag21}) is the same as that in Eq.~(\ref{eq:offdiag12}) if the transpose of the latter is taken, we focus just on the implications of Eq.~(\ref{eq:offdiag12}). Multiplying this equation by $\widehat{\bf s}^{-1}$ and $\widehat{\bf c}^{-1}$ from the left and right, respectively, we get
    \begin{eqnarray}
    \widetilde{\bf m}^{}_{\rm R} = \widehat{\bf s}^{-1}\widehat{\bf c} \widetilde{\bf m}^{}_{\rm D} - \widetilde{\bf m}^{\rm T}_{\rm D} \widehat{\bf s}\widehat{\bf c}^{-1} \; .
    \label{eq:mRtil}
    \end{eqnarray}
    By using the relation in Eq.~(\ref{eq:DijDji}), one can easily verify that the matrix on the right-hand side of Eq.~(\ref{eq:mRtil}) is indeed symmetric. Hence the requirement for the matrices on both sides of Eq.~(\ref{eq:mRtil}) to be symmetric doesn't give more information. However, Eq.~(\ref{eq:mRtil}) implies
    \begin{eqnarray}
    \left( \widetilde{\bf m}^{}_{\rm R} \right)^{}_{ij} = \frac{c^2_i - s^2_j}{s^{}_i c^{}_i} \left( \widetilde{\bf m}^{}_{\rm D} \right)^{}_{ij} \; ,
    \label{eq:mRmD}
    \end{eqnarray}
    which sets a very strong correlation between the elements of those two matrices. As indicated by Eq.~(\ref{eq:mRmD}), the phases of the matrix elements of $\widetilde{\bf m}^{}_{\rm R}$ exactly match those of $\widetilde{\bf m}^{}_{\rm D}$, while the absolute values of the matrix elements of $\widetilde{\bf m}^{}_{\rm R}$ are proportional to those of $\widetilde{\bf m}^{}_{\rm D}$. When taking the diagonal elements, i.e., setting $i=j$ on both sides of Eq.~(\ref{eq:mRmD}), we can figure out three rotation angles
    \begin{eqnarray}
    \frac{s^{}_i c^{}_i}{c^2_i - s^2_i} = \left( \widetilde{\bf m}^{}_{\rm D} \right)^{}_{ii} \left( \widetilde{\bf m}^{}_{\rm R} \right)^{-1}_{ii} \; ,
    \label{eq:tgcs}
    \end{eqnarray}
    or equivalently
    \begin{eqnarray}
    s^2_i = \frac{1}{2} \left[ 1 - \frac{\left( \widetilde{\bf m}^{}_{\rm R} \right)^{}_{ii}}{\sqrt{\left( \widetilde{\bf m}^{}_{\rm R} \right)^{2}_{ii} + 4 \left( \widetilde{\bf m}^{}_{\rm D} \right)^{2}_{ii}}}\right] \; , \quad c^2_i = \frac{1}{2} \left[ 1 + \frac{\left( \widetilde{\bf m}^{}_{\rm R} \right)^{}_{ii}}{\sqrt{\left( \widetilde{\bf m}^{}_{\rm R} \right)^{2}_{ii} + 4 \left( \widetilde{\bf m}^{}_{\rm D} \right)^{2}_{ii}}}\right] \; .
    \label{eq:sc}
    \end{eqnarray}
Notice that all these results obtained thus far are exact without any approximations. Unfortunately, the determination of ${\bf u}^{}_1$ and ${\bf u}^{}_2$ from Eq.~(\ref{eq:mRtil}) is only implicit, which can be recast into
\begin{eqnarray}
{\bf m}^{}_{\rm R} = {\bf u}^{}_2 \widehat{\bf s}^{-1}\widehat{\bf c} {\bf u}^\dagger_1 {\bf m}^{}_{\rm D} - {\bf m}^{\rm T}_{\rm D} {\bf u}^*_1 \widehat{\bf s}\widehat{\bf c}^{-1} {\bf u}^{\rm T}_2 \; ,
\label{eq:mR}
\end{eqnarray}
where the original mass matrices ${\bf m}^{}_{\rm D}$ and ${\bf m}^{}_{\rm R}$ have been recovered.
\end{itemize}

Given ${\bf m}^{}_{\rm D}$ and ${\bf m}^{}_{\rm R}$ in the canonical seesaw model, we expect that $\widetilde{\bf m}^{}_{\bf D} \equiv {\bf u}^\dagger_1 {\bf m}^{}_{\rm D} {\bf u}^*_2$ and $\widetilde{\bf m}^{}_{\rm R} \equiv {\bf u}^\dagger_2 {\bf m}^{}_{\rm R} {\bf u}^*_2$ are on the same order of ${\bf m}^{}_{\rm D}$ and ${\bf m}^{}_{\rm R}$, respectively, because ${\bf u}^{}_1$ and ${\bf u}^{}_2$ are unitary matrices. Thus the results in Eq.~(\ref{eq:tgcs}) and Eq.~(\ref{eq:sc}) readily imply that $s^{}_i \approx {\cal O}(v/M^{}_i)$ is small, as ${\cal O}({\bf m}^{}_{\rm D}) \sim v$ and ${\cal O}({\bf m}^{}_{\rm R}) \sim M^{}_i \gg v$ are usually assumed in realistic seesaw models. In the leading-order approximation, one can neglect the second term proportional to $\widehat{\bf s}$ on the right-hand side of Eq.~(\ref{eq:mR}) and then derive
\begin{eqnarray}
{\bf u}^{}_1 \widehat{\bf c}^{-1} \widehat{\bf s} {\bf u}^\dagger_2 \approx {\bf m}^{}_{\rm D} {\bf m}^{-1}_{\rm R} \; .
\label{eq:approx}
\end{eqnarray}
This indicates that the singular value decomposition of ${\bf m}^{}_{\rm D} {\bf m}^{-1}_{\rm R}$ gives a unique solution of ${\bf u}^{}_1$, ${\bf u}^{}_2$ and $\widehat{\bf s}$ (or $\widehat{\bf c}$). With these matrices, we can further implement Eqs.~(\ref{eq:mLeffsim}) and (\ref{eq:mReffsim}) to determine ${\bf v}^{}_1$ and ${\bf v}^{}_2$ together with $\widehat{\bf m}$ and $\widehat{\bf M}$, respectively.

In order to confront the seesaw model with future precision measurements and extract the model parameters, one must in the first place adopt an explicit parametrization of the seesaw model itself. Such a parametrization can be performed in different but equivalent ways. In the last part of this section, we shall clarify the physical parameters to be taken in the Wigner parametrization, since not all the involved parameters in Eq.~(\ref{eq:wigner}) are relevant. Moreover, the connection of this parametrization to others in the literature will be established.

\underline{\it Generic Parametrization} --- As indicated by the Lagrangian in Eq.~(\ref{eq:Lag}), additional free parameters in the canonical seesaw model, when compared to the SM, arise from the Yukawa coupling matrix ${\bf y}^{}_\nu$ and the Majorana mass matrix ${\bf m}^{}_{\rm R}$ of three right-handed neutrinos. However, not all those parameters are physical due to the freedom of choosing a flavor basis. In the flavor basis where both the charged-lepton mass matrix and right-handed neutrino mass matrix are diagonal, we have three eigenvalues from $\widehat{\bf m}^{}_{\rm R} = {\rm diag}\{r^{}_1, r^{}_2, r^{}_3\}$ and fifteen parameters from the Yukawa coupling matrix $\widetilde{\bf y}^{}_\nu$, which is generally an arbitrary $3\times 3$ complex matrix but three phases of it can be absorbed into the charged-lepton fields. In this case, we can simply parametrize the seesaw model in terms of eighteen real parameters
    \begin{eqnarray}
    \widetilde{\bf y}^{}_\nu = \left(\begin{matrix} |y^{}_{11}| & |y^{}_{12}| e^{{\rm i}\varphi^{}_{12}} & |y^{}_{13}| e^{{\rm i}\varphi^{}_{13}} \cr |y^{}_{21}| & |y^{}_{22}| e^{{\rm i}\varphi^{}_{22}} & |y^{}_{23}| e^{{\rm i}\varphi^{}_{23}} \cr |y^{}_{31}| & |y^{}_{32}| e^{{\rm i}\varphi^{}_{32}} & |y^{}_{33}| e^{{\rm i}\varphi^{}_{33}}\end{matrix}\right) \; , \quad \widehat{\bf m}^{}_{\rm R} = \left( \begin{matrix} r^{}_1 & 0 & 0 \cr 0 & r^{}_2 & 0 \cr 0 & 0 & r^{}_3 \end{matrix} \right) \; ,
    \label{eq:ynur}
    \end{eqnarray}
where $r^{}_i$ (for $i = 1, 2, 3$) and $|y^{}_{ij}|$ (for $i, j = 1, 2, 3$) are non-negative real numbers and $\varphi^{}_{ij}$ (for $i = 1, 2, 3$ and $j = 2, 3$) are arbitrary phases. This parametrization involves the generic parameters in the Lagrangian, but it is unclear how they are related to physical observables at low energies, such as nonzero masses of light Majorana neutrinos, lepton flavor mixing angles and CP-violating phases.

\underline{\it Euler Parametrization} --- As has been shown in Eq.~(\ref{eq:diag}), the $6\times 6$ neutrino mass matrix can be diagonalized by a $6\times 6$ unitary matrix via ${\cal U}^\dagger {\cal M} {\cal U}^* = \widehat{\cal M}$, where we have defined
\begin{eqnarray}
{\cal M} \equiv \left( \begin{matrix} {\bf 0} & {\bf m}^{}_{\rm D} \cr {\bf m}^{\rm T}_{\rm D} & {\bf m}^{}_{\rm R} \end{matrix} \right) \; , \quad \widehat{\cal M} \equiv \left( \begin{matrix} \widehat{\bf m} & {\bf 0} \cr {\bf 0} & \widehat{\bf M} \end{matrix} \right) \; . \label{eq:MUM}
\end{eqnarray}
Notice that ${\bf m}^{}_{\rm D} = {\bf y}^{}_\nu v/\sqrt{2}$ and ${\bf m}^{}_{\rm R}$, as mentioned before, contain all model parameters, but some of them are unphysical. In general, ${\bf m}^{}_{\rm D} = {\bf y}^{}_\nu v/\sqrt{2}$ is supposed to be an arbitrary $3\times 3$ complex matrix, involving eighteen real parameters; while ${\bf m}^{}_{\rm R}$ is a $3\times 3$ complex and symmetric matrix, having twelve real parameters. According to Eq.~(\ref{eq:diag}), these thirty parameters can be expressed in terms of six eigenvalues in $\widehat{\cal M}$ and thirty-six parameters from the $6\times 6$ unitary matrix ${\cal U}$, i.e.,
\begin{eqnarray}
\left( \begin{matrix} {\bf 0} & {\bf m}^{}_{\rm D} \cr {\bf m}^{\rm T}_{\rm D} & {\bf m}^{}_{\rm R} \end{matrix} \right) = \left(\begin{matrix} {\bf V} & {\bf R} \cr {\bf S} & {\bf U}\end{matrix}\right) \cdot \left( \begin{matrix} \widehat{\bf m} & {\bf 0} \cr {\bf 0} & \widehat{\bf M} \end{matrix} \right) \cdot \left(\begin{matrix} {\bf V} & {\bf R} \cr {\bf S} & {\bf U}\end{matrix}\right)^{\rm T} \; ,\label{eq:MeqU}
\end{eqnarray}
where the exact matching between two sets of parameters on both sides is actually guaranteed by the following identity
\begin{eqnarray}
{\bf V} \widehat{\bf m} {\bf V}^{\rm T} + {\bf R} \widehat{\bf M} {\bf R}^{\rm T} = {\bf 0} \; .
\label{eq:seesaw}
\end{eqnarray}
It is straightforward to see that twelve real parameters can in fact be removed from the theory according to the constraints in Eq.~(\ref{eq:seesaw}).

To fully parametrize the canonical seesaw model by using physical masses of Majorana neutrinos, flavor mixing angles and CP-violating phases, Xing has advocated the Euler-like parametrization of the $6\times 6$ unitary matrix ${\cal U}$ in a factorized form~\cite{Xing:2011ur, Xing:2020ijf}
\begin{eqnarray}
{\cal U} = \left( \begin{matrix} {\bf 1} & {\bf 0} \cr {\bf 0} & {\bf U}^{}_0\end{matrix} \right) \cdot \left( \begin{matrix} {\bf A} & {\bf R} \cr {\bf D} & {\bf B}\end{matrix} \right) \cdot \left( \begin{matrix} {\bf V}^{}_0 & {\bf 0} \cr {\bf 0} & {\bf 1}\end{matrix}\right) \; ,\label{eq:xing}
\end{eqnarray}
where the Euler-like parametrizations of those three $6\times 6$ unitary matrices on the right-hand side are given by
\begin{eqnarray}
\left( \begin{matrix} {\bf 1} & {\bf 0} \cr {\bf 0} & {\bf U}^{}_0\end{matrix} \right) &=& {\cal O}^{}_{56}\cdot {\cal O}^{}_{46} \cdot {\cal O}^{}_{45} \; , \\
\left( \begin{matrix} {\bf V}^{}_0 & {\bf 0} \cr {\bf 0} & {\bf 1} \end{matrix} \right) &=& {\cal O}^{}_{23}\cdot {\cal O}^{}_{13} \cdot {\cal O}^{}_{12} \; , \\
\left( \begin{matrix} {\bf A} & {\bf R} \cr {\bf D} & {\bf B} \end{matrix} \right) &=& {\cal O}^{}_{36}\cdot {\cal O}^{}_{26} \cdot {\cal O}^{}_{16} \cdot {\cal O}^{}_{35} \cdot {\cal O}^{}_{25} \cdot {\cal O}^{}_{15} \cdot {\cal O}^{}_{34} \cdot {\cal O}^{}_{24} \cdot {\cal O}^{}_{14}\; .
\end{eqnarray}
The above fifteen $6\times 6$ unitary matrices ${\cal O}^{}_{ij}$ (for $1 \leq i<j \leq 6$) can be constructed by setting the diagonal $(i, i)$- and $(j, j)$-elements to $c^{}_{ij} \equiv \cos \theta^{}_{ij}$ while other diagonal elements to one, and the off-diagonal $(i,j)$- and $(j,i)$-elements to $\tilde{s}^{*}_{ij} \equiv e^{-{\rm i}\delta^{}_{ij}} \sin \theta^{}_{ij}$ and $-\tilde{s}^{}_{ij} \equiv -e^{{\rm i}\delta^{}_{ij}} \sin\theta^{}_{ij}$, respectively, while other off-diagonal elements to zero. There are only fifteen rotation angles $\theta^{}_{ij}$ and fifteen CP-violating phases $\delta^{}_{ij}$ (for $1 \leq i < j \leq 6$) in the $6\times 6$ unitary matrix ${\cal U}$, which should possess thirty-six parameters.\footnote{For a general $n\times n$ unitary matrix ${\cal V}$, its Euler parametrization is written as ${\cal V} = {\cal P}\cdot {\cal U}$, where ${\cal U}$ can be decomposed into the product of a series of two-dimensional complex rotations involving $n(n-1)/2$ rotation angles and $n(n-1)/2$ phases. As the number of non-factorizable phases is $(n-1)(n-2)/2$, there are already $n(n-1)/2 - (n-1)(n-2)/2 = n - 1$ factorizable phases in ${\cal U}$. Therefore, ${\cal P}$ stands for a diagonal matrix with $n$ phases such that the total number of real parameters in ${\cal U}$ is $n(n-1)/2 + n(n-1)/2 + n = n^2$.} The missing phases can be added back by multiplying ${\cal U}$ with a diagonal phase matrix ${\cal P} \equiv {\rm diag}\{e^{{\rm i}\phi^{}_1}, e^{{\rm i}\phi^{}_2}, e^{{\rm i}\phi^{}_3}, e^{{\rm i}\phi^{}_4}, e^{{\rm i}\phi^{}_5}, e^{{\rm i}\phi^{}_6}\}$ from the left. Consequently, we obtain the Euler parametrization of a general $6\times 6$ unitary matrix with thirty-six real parameters.

Comparing the explicit form of ${\cal U}$ in Eq.~(\ref{eq:wigner}) and that in Eq.~(\ref{eq:xing}), one can immediately recognize ${\bf V} = {\bf A}\cdot {\bf V}^{}_0$, ${\bf S} = {\bf U}^{}_0 \cdot {\bf D} \cdot {\bf V}^{}_0$ and ${\bf U} = {\bf U}^{}_0 \cdot {\bf B}$, while ${\bf R}$ remains unchanged. Since only ${\bf V}$ and ${\bf R}$ appear in the interactions and three unphysical phases can be absorbed by redefining the charged-lepton fields, implying that the phase matrix ${\cal P}$ and the parameters in ${\bf U}^{}_0$ are irrelevant, we are left with twelve rotation angles $\theta^{}_{ij}$, twelve CP phases $\delta^{}_{ij}$ (for $1 \leq i \leq 3$ and $i<j\leq 6$) and six mass eigenvalues from $\widehat{\bf m}$ and $\widehat{\bf M}$. From the identity in Eq.~(\ref{eq:seesaw}), one can remove twelve parameters by expressing them in terms of the remaining eighteen independent parameters. As demonstrated by Xing in a series of works~\cite{Xing:2023adc, Xing:2023kdj, Xing:2024xwb, Xing:2024gmy}, it is physically appealing to choose $\{m^{}_1, m^{}_2, m^{}_3\}$, $\{\theta^{}_{12}, \theta^{}_{13}, \theta^{}_{23}\}$, $\{\delta^{}_{12}, \delta^{}_{13}, \delta^{}_{23}\}$ and $\{\delta^{}_{14}, \delta^{}_{24}, \delta^{}_{34}\}$ as derivational parameters and take the rest eighteen model parameters as original: three heavy Majorana neutrino masses $\{M^{}_1, M^{}_2, M^{}_3\}$, nine rotation angles $\theta^{}_{ij}$ for $i = 1, 2, 3$ and $j = 4, 5, 6$, and six CP-violating phases $\{\delta^{}_{i5} - \delta^{}_{i4}, \delta^{}_{i6} - \delta^{}_{i4}\}$ for $i = 1, 2, 3$.

It will be helpful to make a direct connection between the Euler parametrization and the generic one. Since the diagonalization of the overall neutrino mass matrix ${\cal M}$ is exact in the former case, one cannot simply identify $\widehat{\bf M}$ with $\widehat{\bf m}^{}_{\rm R}$. However, at the leading order of ${\cal O}(v/M^{}_i)$, we can find $\widehat{\bf m}^{}_{\rm R} \approx \widehat{\bf M}$ and $\widetilde{\bf y}^{}_\nu \approx \sqrt{2} {\bf A}^{-1} {\bf R} \widehat{\bf M}/v$, from which one can establish one-to-one correspondence between the parameters in these two parametrizations. Beyond the leading order, one must take into account higher-order corrections of ${\cal O}(v^3/M^3_i)$, and thus the correspondence between two sets of parameters is no longer simple.

\underline{\it Wigner Parametrization} --- First of all, the parameter counting is necessary. In Eq.~(\ref{eq:wigner}), we have four $3\times 3$ unitary matrices $\{{\bf u}^{}_1, {\bf u}^{}_2\}$ and $\{{\bf v}^{}_1, {\bf v}^{}_2\}$, involving $4\cdot 3^2 = 36$ real parameters. It is possible to eliminate three common phases in three rows of either ${\bf v}^{}_1$ or ${\bf v}^{}_2$, but we need three rotation angles $\vartheta^{}_i$ (for $i = 1, 2, 3$) from $\widehat{\bf s}$ or $\widehat{\bf c}$. In consideration of six mass eigenvalues in $\widehat{\bf m}$ and $\widehat{\bf M}$, we have totally $36 + 6 = 42$ real parameters, which should be true for a general $6\times 6$ unitary matrix ${\cal U}$ and a diagonal $6\times 6$ matrix $\widehat{\cal M}$. However, the overall neutrino mass matrix ${\cal M} = {\cal U} \widehat{\cal M} {\cal U}^{\rm T}$ has a zero $3\times 3$ block, leading to the seesaw relation in Eq.~(\ref{eq:seesaw}). In the Wigner parametrization, it is equivalent to
\begin{eqnarray}
    {\bf v}^{}_1 \widehat{\bf m} {\bf v}^{\rm T}_1 = - \widehat{\bf s} \widehat{\bf c}^{-1} \cdot {\bf v}^{}_2 \widehat{\bf M} {\bf v}^{\rm T}_2 \cdot \widehat{\bf s} \widehat{\bf c}^{-1} \; ,
\label{eq:seesawwig}
\end{eqnarray}
with the help of Eqs.~(\ref{eq:mLeffsim}) and (\ref{eq:mReffsim}). This relation is suggestive of physical parameters in the Wigner parametrization: three rotation angles $\vartheta^{}_i$ from $\widehat{\bf s}$ (or $\widehat{\bf c}$), three heavy Majorana neutrino masses $M^{}_i$ from $\widehat{\bf M}$, six parameters from ${\bf v}^{}_2$ (including three rotation angles and three CP-violating phases) and six parameters from ${\bf u}^{}_1$ (including another three rotation angles and three CP-violating phases). These eighteen parameters are physical, which should be compared with those $\{|y^{}_{ij}|, \varphi^{}_{ij}, r^{}_i\}$ in Eq.~(\ref{eq:ynur}) for the generic parametrization. Note that each of the $3\times 3$ unitary matrices ${\bf u}^{}_1$ and ${\bf v}^{}_2$ generally involves additional three CP-violating phases, which have been absorbed by redefining the charged-lepton fields in the former case and chosen to be vanishing in the latter by our convention. To be more explicit, we choose the physical parameters in ${\bf u}^{}_1$ as below
\begin{eqnarray}
{\bf u}^{}_1 = \left( \begin{matrix} c^{}_{\alpha^{}_1} c^{}_{\gamma^{}_1} & s^{}_{\alpha^{}_1} c^{}_{\gamma^{}_1} & s^{}_{\gamma^{}_1} e^{-{\rm i}\delta^{}_1} \cr -s^{}_{\alpha^{}_1} c^{}_{\sigma^{}_1} - c^{}_{\alpha^{}_1} s^{}_{\beta^{}_1} s^{}_{\gamma^{}_1} e^{{\rm i}\delta^{}_1} & c^{}_{\alpha^{}_1} c^{}_{\beta^{}_1} - s^{}_{\alpha^{}_1} s^{}_{\beta^{}_1} s^{}_{\gamma^{}_1} e^{{\rm i}\delta^{}_1} & s^{}_{\beta^{}_1} c^{}_{\gamma^{}_1} \cr s^{}_{\alpha^{}_1} s^{}_{\beta^{}_1} - c^{}_{\alpha^{}_1} c^{}_{\beta^{}_1} s^{}_{\gamma^{}_1} e^{{\rm i}\delta^{}_1} & -c^{}_{\alpha^{}_1} s^{}_{\beta^{}_1} - s^{}_{\alpha^{}_1} c^{}_{\beta^{}_1} s^{}_{\gamma^{}_1} e^{{\rm i}\delta^{}_1} & c^{}_{\beta^{}_1} c^{}_{\gamma^{}_1}\end{matrix} \right) \cdot \left( \begin{matrix} e^{{\rm i}\varphi^{}_1} & 0 & 0 \cr 0 & e^{{\rm i}\phi^{}_1} & 0 \cr 0 & 0 & 1 \end{matrix} \right) \; ,
\label{eq:u1para}
\end{eqnarray}
with $s^{}_{\alpha^{}_1} \equiv \sin\alpha^{}_1$ and $c^{}_{\alpha^{}_1} \equiv \cos\alpha^{}_1$ have been defined and likewise for other rotation angles $\{\beta^{}_1, \gamma^{}_1\}$. For ${\bf v}^{}_2$, we can replace $\{\alpha^{}_1, \beta^{}_1, \gamma^{}_1\}$ and $\{\delta^{}_1, \varphi^{}_1, \phi^{}_1\}$ in Eq.~(\ref{eq:u1para}) by $\{\alpha^{}_2, \beta^{}_2, \gamma^{}_2\}$ and $\{\delta^{}_2, \varphi^{}_2, \phi^{}_2\}$. The remaining six parameters are three rotation angles $\{\vartheta^{}_1, \vartheta^{}_2, \vartheta^{}_3\}$ and heavy Majorana neutrino masses $\{M^{}_1, M^{}_2, M^{}_3\}$.

Let us relate the Wigner parametrization to the Euler one. The heavy Majorana neutrino masses $\{M^{}_1, M^{}_2, M^{}_3\}$ are common physical parameters in both cases. Then the connection is quite transparent
\begin{eqnarray}
{\bf V} = {\bf A} \cdot {\bf V}^{}_0 = {\bf u}^{}_1 \cdot \widehat{\bf c} \cdot {\bf v}^{}_1\; , \quad {\bf R} = {\bf u}^{}_1 \cdot \widehat{\bf s} \cdot {\bf v}^{}_2 \; ,
\label{eq:EulerWigner}
\end{eqnarray}
where ${\bf v}^{}_1$ is determined from the seesaw relation in Eq.~(\ref{eq:seesawwig}) for the Wigner parametrization. It is evident that all the fifteen physical parameters, except for three heavy Majorana neutrino masses, residing in ${\bf R}$ in the Euler parametrization are now converted into its singular values in $\widehat{\bf s}$ and two unitary matrices ${\bf u}^{}_1$ and ${\bf v}^{}_2$ in the Wigner parametrization via the singular value decomposition.

\section{Summary}\label{sec:summ}

In this work, we apply the Wigner parametrization of general unitary matrices to the seesaw model and make several interesting observations. First, it is possible to transform the Dirac neutrino mass matrix ${\bf m}^{}_{\rm D}$ and the Majorana mass matrix ${\bf m}^{}_{\rm R}$ in the flavor space of lepton doublets and right-handed neutrinos, namely, ${\bf m}^{}_{\rm D} \to \widetilde{\bf m}^{}_{\rm D} = {\bf u}^\dagger_1 {\bf m}^{}_{\rm D} {\bf u}^*_2$ and ${\bf m}^{}_{\rm R} \to \widetilde{\bf m}^{}_{\rm R} = {\bf u}^\dagger_2 {\bf m}^{}_{\rm R} {\bf u}^*_2$, such that the resultant overall neutrino mass matrix can be block-diagonalized by an orthogonal matrix with three rotation angles $\{\vartheta^{}_1, \vartheta^{}_2, \vartheta^{}_3\}$. These angles capture the strong hierarchy between the electroweak scale $\Lambda^{}_{\rm EW} \approx 10^{2}~{\rm GeV}$ and the seesaw scale $\Lambda^{}_{\rm SS} \approx 10^{14}~{\rm GeV}$, i.e., $\vartheta^{}_i \sim {\cal O}(\Lambda^{}_{\rm EW}/\Lambda^{}_{\rm SS}) \ll 1$. Second, the effective mass matrix of light Majorana neutrinos in this flavor basis is related to that of heavy Majorana neutrinos by a scaling law $\left({\bf v}^{}_1 \widehat{\bf m} {\bf v}^{\rm T}_1\right)^{}_{ij} = - \left({\bf v}^{}_2 \widehat{\bf M} {\bf v}^{\rm T}_2\right)^{}_{ij} \tan\vartheta^{}_i \tan\vartheta^{}_j$. Therefore, the flavor structure of ${\bf m}^{}_{\rm D}$ and that of ${\bf m}^{}_{\rm R}$ characterized by unitary matrices in the Wigner parametrization have been disentangled from the strong hierarchy among their eigenvalues. The latter has already been incorporated into the rotation angles $\vartheta^{}_i$.

Although all different parametrizations of the seesaw model should be mathematically equivalent, one of them may be more advantageous physically than another in the treatment of specific problems. We have also made a connection of the Wigner parametrization to those existing in the literature and suggested a set of eighteen physical parameters, which can be implemented to fully describe the canonical seesaw model. The extension of such a parametrization to other neutrino mass models and the exploration of its phenomenological implications are also interesting. These issues will be left for future works.

\section*{Acknowledgments}

The author is indebted to Prof. Zhi-zhong Xing for valuable discussions and encouraging comments. This work was supported in part by the National Natural Science Foundation of China under grant No. 12475113 and by the CAS Project for Young Scientists in Basic Research (YSBR-099).

\end{document}